\documentclass[fleqn,10pt]{wlscirep}
\usepackage[utf8]{inputenc}
\usepackage[T1]{fontenc}                              

\title{A reaction-diffusion model for describing the ring/gap structure in disks surrounding individual young stars}

\author[1,*]{Enrique Lopez-Cabarcos}

\affil[1]{Department of Chemistry in Pharmaceutical Sciences, Complutense University of Madrid, Madrid 28040, Spain}

\affil[*]{cabarcos@ucm.es}


\begin{abstract}
The embedded disks surrounding individual Class 0 protostars  are structureless; disks surrounding Class I stars may be continuous or have a ring/gap substructure, whereas all disks around Class II stars have a ring/gap substructure that gradually disappear as the disks evolve into debris disks. This common sequence in young lone stars requires an explanation. This study aims to show that the physical model "Reaction-Diffusion Systems with Moving Reaction Front" (RDS-MRF) can be used to describe and classify protostellar disks according to their structure. A comprehensive review of observations made with the ALMA radio telescope shows: i) that the protostar-disk system presents a geometry analogous to that of an RDS-MRF system with two separate compartments: protostar and disk; ii) that in the protostar, matter is processed at high temperature, resulting in a chemical composition different from that of the disk; and iii) that the equatorial outflow emitted by the protostar, rich in highly reactive H$_3^+$ ions, acts as an MRF that triggers the formation of molecules and nuclei in the disk. The time lag of nucleation with respect to the passage of the MRF would be the cause of the formation of the gaps between the rings of particles that form in the disk. The MRF is a transient phenomenon and its passage causes the transformation of a continuous disk (Class 0) into a disk with a ring/gap structure (Class II) whose temporal evolution begins at the star's interface and moves outwards.

\end{abstract}
\begin{document}

\flushbottom
\maketitle
\section{Introduction}
Observation of dust and gas disks around young stars began at the end of the last century \cite{smith1984circumstellar}. More recently, high-resolution images from the Atacama Large Millimeter/submillimeter Array (ALMA) radio telescope have revealed a wide variety of substructures in these disks of gas and dust, and there is no longer any doubt that this is the place where planets are formed \cite{sallum2015accreting, muller2018orbital,haffert2019two}. Among the programs carried out at ALMA, the Disk Substructures at High Angular Resolution Project (DSHARP) \cite{huang2018disk,dullemond2018disk} have characterised the spatial distribution of solid particles for a sample of 20 protoplanetary disks at $\sim$ 5 au resolution (at the distance of 125 pc) by means of observations of the continuum emission from micron to millimetre-sized particles. ALMA observations confirmed that in many disks there is an annular structure consisting of particle rings separated by gaps where the particle density is depleted \cite{isella2016ringed,cieza2017alma,fedele2018alma,ansdell2018alma,van2019protoplanetary}. ALMA also provides information on the molecular emission of $^{12}$CO and isotopologues of CO, which are associated with the distribution of gas in the disks. Thus, the observation of the $^{12}$CO (J = 2-1) emission line corresponding to the disks surrounding the young stars HD 163296 \cite{isella2016ringed} and AS 209 \cite{guzman2018disk} was performed with a resolution $\sim$ 10 au. 

The chemical composition of both the protostar and the disk is less known due to the lack of resolution to specify the variation of the composition with distance, especially near the protostar. However, the situation has improved somewhat with the publication of the results obtained by ALMA with the "Molecules at Planetary Formation Scales" (MAPS) program \cite{oberg2021molecules}. Using spatial resolution between 7 and 30 au, MAPS explores the chemical structures in five disks around IM Lup (age $\sim$ 1 Myr), GM Aur (age $\sim$ 3 to 10 Myr), AS 209 (age $\sim$ 1 to 2 Myr), HD 163296 (age $\sim$ 6 Myr), and MWC 480 (age $\sim$  7 Myr). One of the main achievements of MAPS is to show that chemical substructures also exist throughout the disks and many of them roughly coincide with the dust substructures while others do not. 

Recently, the results of the ALMA program Early Planet Formation in Embedded Disks (eDisk) were presented by Ohashi et al \cite{ohashi2023early,lin2023early,yamato2023early}. The eDisk program focuses on searching for substructures in disks
around 12 Class 0 and 7 Class I protostars through continuum 1.3 mm observations at a resolution of about 7 au (0".04). The results show that continuum emission arising from the dust disks is detected around all 19 targets and has little or no substructure in contrast to what happens in Class II protostars. The C$^{18}$O (2-1) emission is detected towards 19 sources, although Kepler's rotation was identified only in 15 cases. 

 Debris disks (DDs) are observed as a late stage in disk evolution and show what happens to disks $\sim$ 10 Myr after their formation, although exceptions occasionally appear and protoplanetary disks (PTPDs) up to 30 Myr old and DDs as young as 6 Myr old can be found. DDs are characterized by having belts of bodies similar to the Kuiper Belt in our Solar System, which is why they are often called exoKuiper belts.
 The results of the \textbf{A}LMA survey to \textbf{R}esolve  exo\textbf{K}uiper belt \textbf{S}ubstructures (ARKS program) were just published in 2026 \cite{marino2026alma,han2026alma,jankovic2026alma}. ARKS analyzed a sample of 24 exoKuiper belts and found a wide range of structures such as narrow belts, wide and smooth belts, and belts with gaps. \textbf{RE}solved \textbf{A}LMA and \textbf{S}ubmillimeter array \textbf{O}bservations of \textbf{N}earby \textbf{S}tars  (REASONS) \cite{matra2025resolved} expanded the sample of resolved belts to 74.  REASONS found that many belts are broad disks rather than narrow rings with fractional widths ($\Delta R/R$) much greater than the Kuiper Belt. This surprising result seems to indicate that planetesimals really do form in the exoKuiper belts, whose great width requires new hypotheses to explain their origin such as the migration of proto-planetary rings  \cite{miller2021formation} or dynamical perturbation of rings by massive planets \cite{booth2009history,lestrade2011stripping}. 

Is there any physical model that describes the transformation of a homogeneous and continuous medium, such as a disk of dust and gas, into a heterogeneous medium with particles grouped in bands separated by particle-depleted regions (gaps)? In Colloidal Science there is a model termed Reaction-Diffusion Systems with Moving Reaction Front (RDS-MRF) that allows this transformation. The model requires that initially the system be made up of two regions spatially separated by an interface, and that in each region there are different reagents that, when diffusing through the interface, react to give new molecules that trigger nucleation. Depending on the concentrations of the reactants and their diffusion coefficients, a moving or stationary reaction front develop. In the case of the moving reaction front (MRF), its passage is followed by the synthesis of new molecules that trigger the nucleation and subsequent formation of periodic bands of particles. Kai et al. \cite{kai1982measurements} with  pH measurements observed the motion of the MRF, and with turbidity measurements, the subsequent formation of nuclei (nucleation front), reporting a time lag between the two. This time lag results in a ring structure of particles (which grow where nuclei are present) separated by gaps (that appear in the neighboring region  depleted of particles). The next ring of particles will appear in the system after the formation of the next nucleation front, and thus the formation of the ring/gap structure will continue until the reactants carried by the MRF are exhausted. G{\'a}lfi \& R{\'a}cz.  \cite{galfi1988properties} used scaling laws to describe the properties of the reaction front and those of the precipitates in a process A + B $\rightarrow$  C. When the concentrations of reactants and their diffusion coefficients are similar in both regions a stationary reaction front (SRF) develops. In this case, the particles concentrate in a single band whose width decreases as the reactant concentration increases
\cite{cabarcos1996crossover}.

The star formation process goes through a stage in which two spatially separated parts appear in the system: the protostar and the disk. Although the matter in both parts comes from the same molecular cloud, it has been processed very differently generating different chemicals in both regions. The observation of stellar outflows associated with star formation is a ubiquitous phenomenon \cite{greenhill1998coexisting,whelan2005resolved}, and although most of them are associated with bipolar jets \cite{mcleod2018parsec} in some cases equatorial outflows have also been observed \cite{greenhill1998coexisting,okoda2021faust}. The difference between a jet and an outflow is that a jet is defined as high velocity ($>$ 100 km s$^{-1}$), highly-collimated ionized gas launched along the axis of rotation of young stellar objects, while outflows refer to low velocity ($<$ 20 km s$^{-1}$) molecular gas entrained by jets and/or winds as they push through the material that surrounds the protostar \cite{klaassen2013alma,mcleod2018parsec}. In this way, the geometry and motion of the reagents in the protostar-disk system provides a situation analogous to that required for an RDS-MRF to exist. According to Klaassen et al., for a complete theory of star formation we still need to understand the mechanisms responsible for jet launching and outflow emission in protostars \cite{klaassen2013alma}.

It is reasonable to think that the matter ejected from the protostar through an equatorial outflow in its outward motion encounters the matter of the protostellar disk triggering chemical reactions of the type A$_i$ + B$_i$ $\rightarrow$ C$_i$. The letter A$_i$ labels any reactant (ion, element, radical, molecule, or even radiation) generated within the protostar that is capable of reacting with the gas and dust in the disk termed B$_i$, generating new substances called C$_i$. The formation of C$_i$ would create the instability necessary for the nucleation, aggregation, and subsequent grain formation phenomena to begin in the disk. Although nucleation from the gas phase is a complex problem, the formation of micro-particles in the gas phase is very rapid (a few minutes) if nucleation is initiated by the arrival of an "initiating species" reacting with another gas containing a "growing species" \cite{adachi1992gas,dingilian2021new}. This physical situation, which is the basis of the chemical vapor deposition technique, is analogous to that of equatorial outflows moving through protostellar disks and this contribution highlights this analogy. The laws of the RDS-MRF model can only be used to describe the initial stage of disk evolution, since the colloidal description of the system ceases to be valid when particle accretion begins and gravity takes control of the process.

\section{Methods: Reaction-Diffusion Systems with Moving Reaction Front (RDS-MRF)}

\begin{figure}[ht]
\centering
\includegraphics[width=\linewidth]{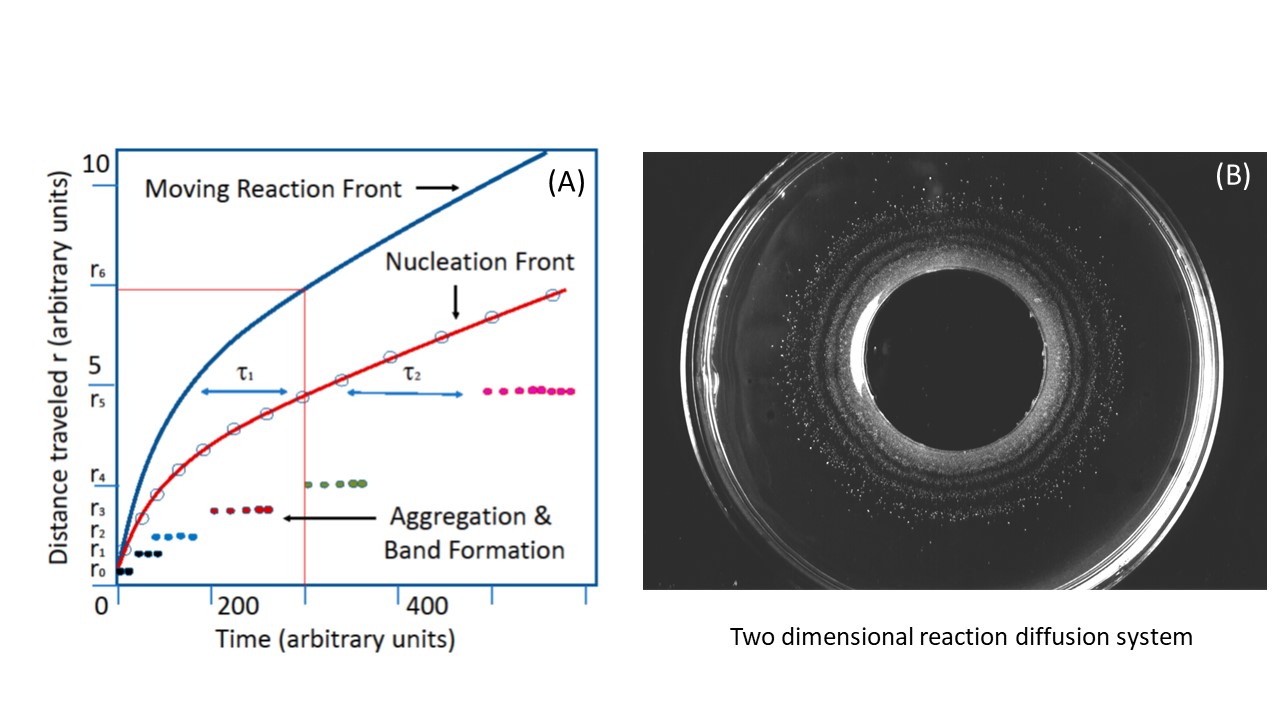}
\caption{A) Distance travelled by the MRF and position of the NF as a function of time. The distance r is measured from the interface between reactants A$_i$ and B$_i$, which is chosen as r$_0=0$, while the time of emission of the MRF is chosen as t$_0=0$. The passage of the reaction front triggers the nucleation and the formation of bands of particles at positions r$_n$, which are represented with coloured dots. The formation of bands is not instantaneous but takes place over a time interval. The time lag between the MRF and the NF is $\tau_1$ whereas the time lag between the NF and the formation of bands is $\tau_2$. (B) Micro particle bands in an RDS-MRF approximating a two-dimensional system. Reagent A (Na$_2$HPO$_4$) is placed within the circular cavity in the centre from where diffuses to find reactant B (CaCl$_2$) that is dissolved in a gel (99.5 per cent $w/w$ is water).}
\label{fig:Nt1-1}
\end{figure}

The essence of the RDS-MRF model could be summarized in three steps. First, for a MRF to exist at least two different reactants are required, spatially separated by an interface in two separate compartments and a force that drives the reactant with the highest chemical potential (A$_i$) to move through the medium where the other (B$_i$) is located. Second, the encounter of both reactants generates C$_i$ molecules (A$_i$ + B$_i$ $\rightarrow$  C$_i$) that trigger the nucleation process, the so-called nucleation front (NF). Third, the nuclei begin to grow, forming particles and aggregates that self-organize into bands, or rings in the case of circular symmetry. Band formation will occur in the region where there are particles and they only appear after nucleation, in other parts of the system where the MRF has not yet arrived, there are no nuclei, particles, or bands (see Figure \ref{fig:Nt1-1}A). 
Fourth, due to the time lag between the moving reaction front and the nucleation front, after the formation of a band there is a depletion of material in the region near it, which leads to the appearance of a gap. Thus, in a medium that was previously continuous, a structure of bands of particles separated by gaps is produced (see Figure \ref{fig:Nt1-1}B). Fifth, the position of the bands, their formation time and their width follow scaling laws.

As illustrated in Figure \ref{fig:Nt1-1}A, at one point in time different phenomena occur in different parts of the system due to the time lags between the MRF and the NF ($\tau_1$), and between the NF and the formation of bands of particles ($\tau_2$). For example, for t = 300 the MRF is reaching the position r$_6$, the nucleation process is already underway at position r$_5$, and at r$_4$  a band of particles begin to appear. At smaller distances r $<$ r$_4$ bands have long since formed and between bands there are gaps since the formation of the band in r$_1$ depletes the matter of the surrounding region until saturation is reached again in r$_2$. Time is the paramount variable in determining what happens at any position within the system. The time delay of the nucleation with respect to the MRF imposes the zoning of the system with regions where the concentration of nuclei reaches the saturation value and the matter begins to aggregate, separated by matter depleted regions. Figure \ref{fig:Nt1-1}B illustrates the formation of bands of particles in an experiment performed in a Petri dish where Na$_2$HPO$_4$ has been placed in the central cavity from where it diffuses to meet CaCl$_2$ embedded in a gel. A band of matter formed at the interface and three bands of particles separated by gaps can be observed. The temporal and spatial sequence of events in RDS-MRF was measured by Kai et al. \cite{kai1982measurements}, who proposed five empirical laws to describe the system. Shortly after, G{\'a}lfi \& R{\'a}cz \cite{galfi1988properties} reported scaling laws in systems with reactions of type A$_i$ + B$_i$ $\rightarrow$ C$_i$.

The position of the MRF, r$_{MRF}$,  and the NF, r$_{NF}$, as a function of time are given by, 
\begin{equation}
    r_{MRF}=K_{1}t^{\alpha}
\label{Eq:Nt1-1}
\end{equation}
\begin{equation}
    r_{NF}=K_{2}t^{\beta}
\label{Eq:Nt1-2}
\end{equation}
The exponents $\alpha$ and $\beta$, and the parameters K$_1$ and K$_2$ must be determined experimentally. The model also provides scaling laws for the position of the center of the bands r$_n$, (Eq. \ref{Eq:Nt1-3}); for the relationship between r$_n$ and the time of formation of the n-band t$_n$, (Eq. \ref{Eq:Nt1-4}); and for the  width of the bands, $\omega_n$ , as a function of their distance from the interface (Eq. \ref{Eq:Nt1-5}).
\begin{equation}
    \frac{r_{n}}{r_{n-1}}\rightarrow p
\label{Eq:Nt1-3}
\end{equation}
\begin{equation}
    r_{n}\sim t_{n}^{1/2}
\label{Eq:Nt1-4}
\end{equation}
\begin{equation}
    \omega_{n}\sim  r_{n}^{\nu}
\label{Eq:Nt1-5}
\end{equation}
 The spacing law (Eq. \ref{Eq:Nt1-3}) states that the ratio between the positions of adjacent bands approaches a constant value p $>$ 1 for n large. The time law (Eq. \ref{Eq:Nt1-4}) relates the position of the n$^{th}$ band to the time t$_n$ of its formation being the exponent $1/2$ characteristic of Fickian diffusion. The width of the bands $\omega_n$ grows with the distance to the interface according to the width law (Eq. \ref{Eq:Nt1-5}) where the exponent $\nu$ is smaller than 1. The MRF is the only front that transports matter (the reactant A$_i$) as it moves through the medium where the reactant B$_i$ is present, while the NF corresponds to a phase change within this medium.
\section{Results}
\subsection{Protostars with disks as RDS-MRF analogues}
Recently observed features in protostar-disk systems are similar to those required in a RDS-MRF.  Namely: 1) The geometry of the system is analogous, the protostar and the disk providing two separate compartments where matter is stored. There are many observations that confirm that individual protostars are surrounded by disks of gas and dust \cite{huang2018disk,dullemond2018disk,ohashi2023early,lin2023early}. 2) Nebular matter is processed in the protostar under conditions of temperature and pressure very different from those of the disk where all the molecules that have been detected are molecules of the interstellar medium (ISM) \cite{mcguire20182018}. Therefore, the two compartments have a different composition: in the protostar, ions predominate, while in the disk, ISM molecules do. The results of the MAPS program support this assumption \cite{oberg2021molecules}. 3) In its outward motion, the equatorial outflow  emitted by the protostar travels through the disk and its atmosphere, behaving as an MRF. Highly reactive molecules such as H$_3^+$ within the equatorial outflow \cite{aikawa2021molecules}, when they encounter ISM molecules (i.e. CO, O, N$_2$; CH$_3$OH, HCN, etc.) in the disk that have barely been modified, produce the new molecules that trigger the nucleation process in the disk. 4) The rings of millimeter/submillimeter particles observed in protostellar disks in the DSHARP program would be the equivalent of the NF in the RDS-MRF model. 5) The model provides a clear explanation for the existence of the disks observed with ALMA: in very young protostars (Class 0 and some Class I) continuous disks were detected with the eDisk program; in Class II protostars disks with annular structure were detected with the DSHARP program, while in disks older than $\sim$ 10 Myr debris belts are observed with the ARKS program. Furthermore, there are indications that the scaling laws of RDS-MRF can also be applied to the particle rings observed in Class II disks.

\begin{figure}[ht]
\centering
\includegraphics[width=12 cm]{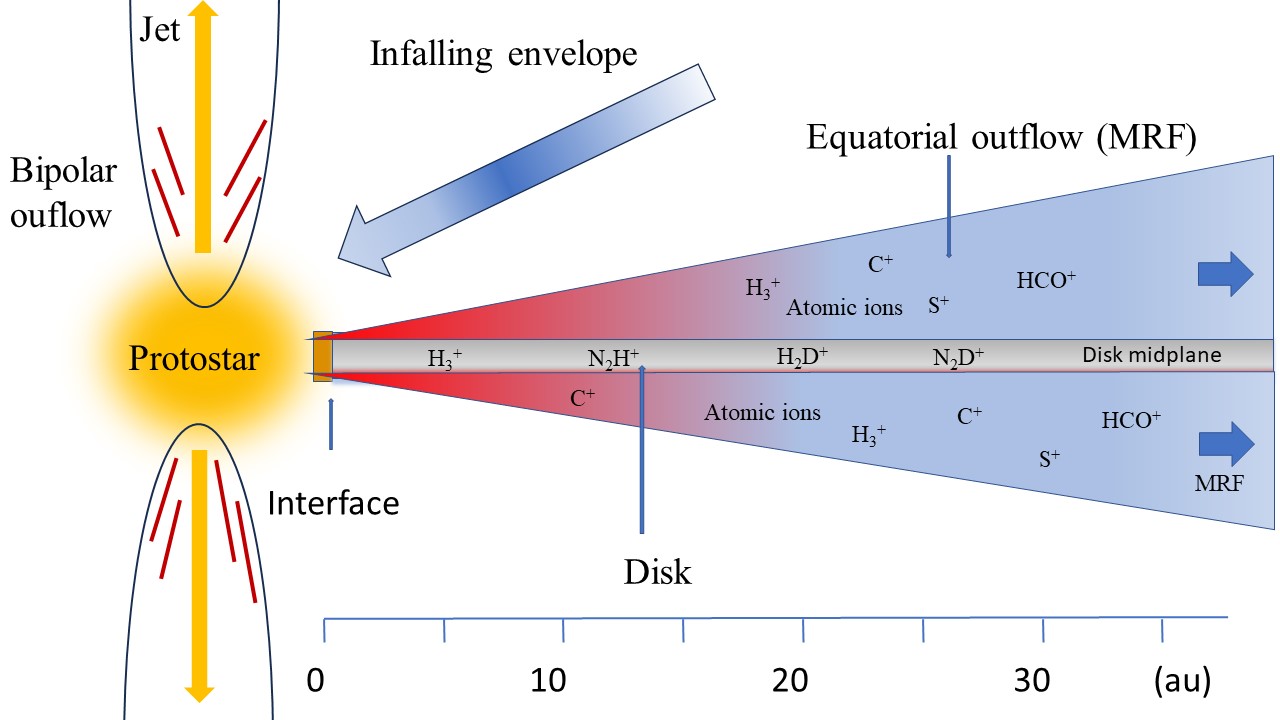}
\caption{Sketch of the RDS-MRF model including a protostar, an interface, an optical thick disk (grey), and an optical thin equatorial outflow (magenta and azure) all viewed edge-on. The interface will be used as the origin of the reference system to measure distances in the disk, The distribution of ions in the disk and its atmosphere is taken from Aikawa et al \cite{aikawa2021molecules}.
}
\label{fig:Nt1-2}
\end{figure}

\subsection{The protostar and the disk are two spatially separated regions}
By definition, a protostar is an object that transforms the gravitational potential energy of matter that collapses towards the center into luminosity. In the early stages of star formation, conservation of angular momentum through gravitational collapse leads to the formation of a disk of gas and dust around the young protostar, thus giving rise to the two components necessary to develop an MRF
\cite{lada1987star}. Star formation involves not only accretion of gas and dust in the center of a nebula but also the expulsion of ionised matter from the protostar in the form of outflows and jets (see Figure \ref{fig:Nt1-2}). According to the slope $\alpha$ of the spectral energy distribution (SED) between 2 and 25 $\mu$m protostars are classified into four classes.  Class 0 in which the final stellar mass has not yet been assembled and only a central condensed core exists; Class I includes protostars with $\alpha > 0 $  and SED broader than single-temperature black body; Class II groups objects with $ -2 < \alpha < 0 $ and SED broader than  single-temperature black body; and Class III contains objects with $\alpha < -2 $  and SED close to that of single-temperature black body. The Class I stage would represent the intermediate stage between the Class 0 phase, dominated by the star´s accretion process and the Class II phase characterized by disks with ring/gap annular substructure.

The protostar is located at the center while the disk is located near the equatorial plane of the protostar (see Figure \ref{fig:Nt1-2}) so the condition of two regions spatially separated by an interface is fulfilled. On the other hand, although the initial fragment of the molecular cloud is almost homogeneous, after dividing into protostar and disk, its matter is processed in two different and separate environments resulting in different compositions in both places.

\begin{figure}[ht]
\centering
\includegraphics[width=\linewidth]{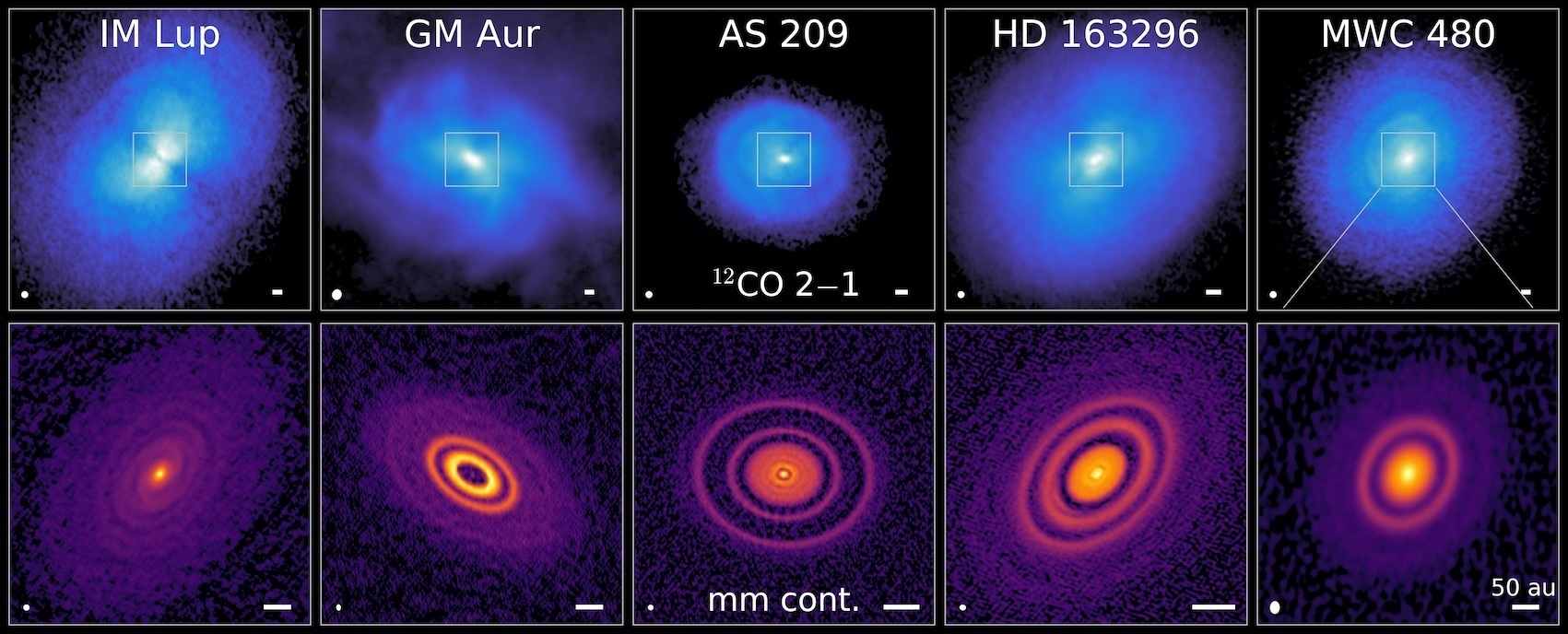}
\caption{Gas and dust disks around young stars. The top row shows zeroth-moment maps of $^{12}$CO (2-1) and the bottom row, the 220 GHz continuous images showing the variety of continuous structures present in the MAPS samples observed at the same spatial resolution. The synthesized beam and a scale bar indicating 50 au are shown in the lower left and right corners, respectively, of each panel. Upper panels are adapted from \cite{oberg2021molecules} and reproduced with {\"O}berg permission. Bottom panels are adapted from \cite{law2021molecules} and reproduced with Law permission. Copyright AAS.}
\label{fig:Nt1-3}
\end{figure}

The two spatially separated regions with different chemical composition have already been observed with the new interferometers in Class 0 \cite{maury2019characterizing}, and Class I protostars \cite{takakuwa2012keplerian}.  The presence of a ring/gap substructure in disks around Class II protostars has also been confirmed \cite{ansdell2018alma,isella2016ringed,huang2018disk,dullemond2018disk}. The results of the eDisk program \cite{ohashi2023early} show that the 12 Class 0 disks studied do not present any type of substructure, while among the Class I disks some do and others do not \cite{yamato2023early, lin2023early}. These results seem to indicate that the disk forms in the Class 0 stage, the formation of substructures begins in the Class I stage and is consolidated in the Class II stage. 

\subsection{The protostar and the disk have different composition}

For an MRF to be generated, the chemical composition of the two regions must be different. The MAPS program has increased the spatial resolution of observations of a considerable number of molecules in five disks around young stars, and it has been recognized that matter is processed in Class 0 protostars \cite{oberg2021molecules,law2021molecules,goicoechea2012complete,bianchi2019astrochemistry}. The consideration of the protostar as a giant chemical reactor rather than a hypothesis is an observable fact that is just beginning to be investigated in detail.

Unlike the matter in the protostar, the matter in the disk at large distances from the star is hardly processed, as shown by the fact that only ISM molecules have so far been found in these regions \cite{mcguire20182018}. To date, at least 204 molecules containing between 2 and 70 atoms have been detected in the ISM. In contrast, the number of molecules detected in protoplanetary disks is only about 24 increasing to 41 if we include isotopologues \cite{mcguire20182018} and five new molecules detected in the disks around MWC 480 and LkCa 15 \cite{loomis2020unbiased}. However, a large fraction of the molecular content of the disks is locked up in the ices located in the mid-plane, and since the detection is based on the rotational transitions occurring in the gas-phase molecules, the amount of detectable emission in the disks is inherently biased. Despite this, we can say that the composition of the protostar is abundant in ionic species, while the molecules of the ISM prevail in the protostellar disk.

Although the $^{12}$CO images in the upper panel of Figure \ref{fig:Nt1-3} show some difference between the younger and older disks, this difference is smaller than that observed in the dust continuum images shown in the bottom panel, where clear difference can be seen between the nearly absence of structure in the IM Lup disk and the ring and gap structure of the other four disks. 

\subsection {Observation of equatorial outflows (MRF analogs)}

Most protostars with mass similar to the Sun go through an outflow phase lasting about 10$^{5}$ yr with mass loss rates up to 10$^{-5}$ M$_{\odot}$ yr$^{-1}$. Given the high resolution achieved with new telescopes in recent decades, the presence of an MRF should be observable, however, there is a possibility that it has been detected but not recognized as such. Jets dragging bipolar outflows emanating from the central core of the collapsing nebula have been observed in young, low- and high-mass stars, and also in brown dwarfs \cite{mcleod2018parsec,whelan2005resolved}, suggesting that there must be a universal mechanism governing jet and outflow emission. However, jets have a bipolar geometry and they cannot travel through the disk; the MRF must be an equatorial outflow. 
While in some cases equatorial outflows have been detected and identified as such \cite{greenhill1998coexisting}, in others they appear as emissions that extend from the source in a direction almost perpendicular to bipolar outflows \cite{okoda2021faust}, and sometimes an alternative explanation is chosen for a phenomenon with equatorial outflow characteristics.

\begin{figure}
    \centering
    \includegraphics[width=11 cm]{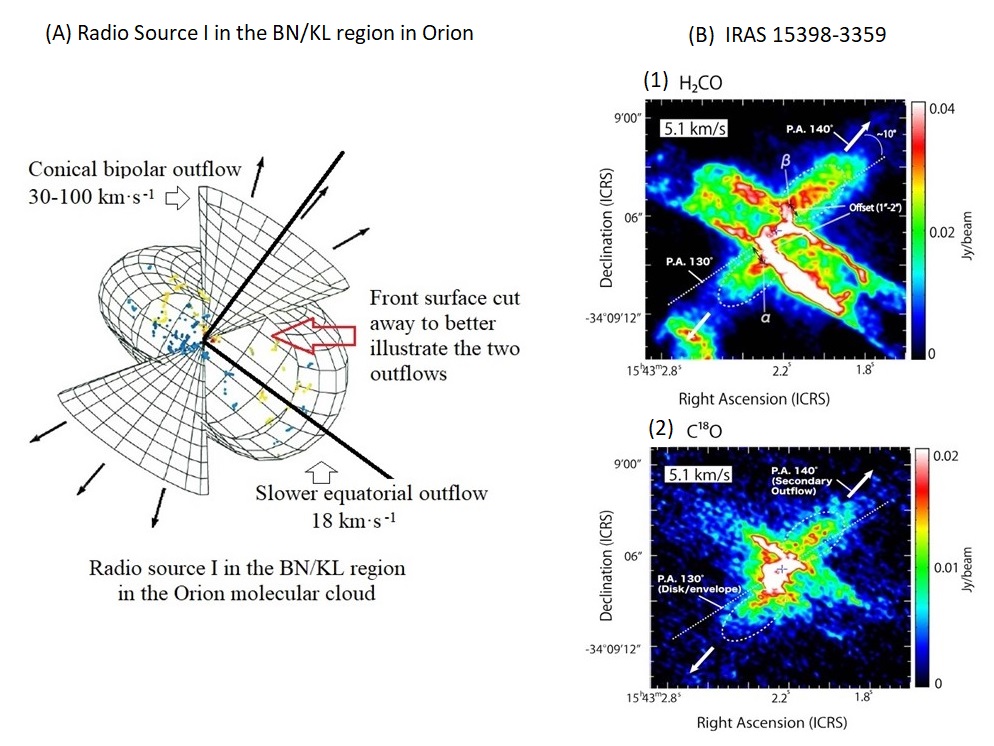}
    \caption{(A) Schematic showing the high-velocity bipolar outflow (between 30 and 100 km s$^{-1}$) represented by two cones and arrows, and the low-speed equatorial outflow (18 km s$^{-1}$) represented by a toroidal surface in the Radio source I system in the BK/KL region in Orion. Left panel is adapted from \cite{greenhill1998coexisting} and is copyright Springer Nature. Reused under license number 5667730238292. (B) Images obtained at ALMA of: (1) H$_2$CO and (2) C$^{18}$O emission lines observed in the Class 0 protostar IRAS 15398-3359. Both emissions show a bipolar outflow extending from northeast to southwest (position angle, P.A. 220°) and an equatorial outflow extending from southeast to northwest (P.A. 130º-140º). The dashed lines represent the disk envelope axis (P.A.130º). Left panel is from \cite{okoda2021faust} and reproducedused under the terms
   of the Creative Commons Attribution 4.0 licence}
    \label{fig:Nt1-4}
\end{figure}

\subsubsection{Equatorial outflow in the Source I protostar in the BN/KL region of the Orion molecular cloud}
The first observation of an equatorial outflow was reported from the so-called Source I protostar in the BN/KL region in the Orion molecular cloud \cite{greenhill1998coexisting}. Greenhill et al., show the coexistence of conical bipolar and toroidal equatorial outflows in the object known as radio source I, which is surrounded by a rotating disk of gas and molecular material. According to their report, a high-velocity (30-100 km $s^{-1}$) conical bipolar outflow is predominant within 60 au of the protostar, while a slower (18 km $s^{-1}$) equatorial outflow orthogonal to the bipolar one extends up to a radius of 1000 au (see Figure \ref{fig:Nt1-4} A). This work shows that outward matter emission from protostars not only occur throughout bipolar outflows but also as an outflow of mass slowly moving outwards along the mid-plane of the disk. In this way, the composition of the disk would be determined not only by the gas and dust accreted from the natal nebula, but also by the chemical processes triggered locally by the passage of the equatorial outflow. 

\subsubsection{Equatorial outflows in the protostars IRAS 15398-3359 and IRAS 04302+2247 (hereafter IRAS 1538 and 04302)}
\begin{itemize}
    \item IRAS 15398 is a class 0 protostar located in the Lupus 1 molecular cloud, which is in the earliest stage of its evolution, and whose mass is estimated to be less than 0.09$M_{\odot}$. Oya et al.\cite{oya2014substellar} reported a bipolar outflow driven from the protostar in the H$_2$CO emission line, which was confirmed by Bjerkeli et al.\cite{bjerkeli2016young} showing that the CO, HCO$^+$ and N$_2$H$^+$ emission lines also trace the motion of the bipolar outflow. In the framework of the FAUST program (\textbf{F}ifty \textbf{AU} \textbf{ST}udy of the chemistry in the disk/envelope system of solar-like protostars) carried out in ALMA by Okoda et al.,\cite{okoda2021faust} a visible secondary outflow was discovered in the emission lines of H$_2$CO (top panel in Figure \ref{fig:Nt1-4} B), SO and C$^{18}$O (bottom panel in Figure \ref{fig:Nt1-4} B) that extends from the source in an almost perpendicular direction (equatorial outflow) to the known bipolar outflow. 
    \item The eDisk program observed Class 0 and Class I protostars at high resolution and therefore should have observed equatorial outflows in some of the 19 protostars investigated. Indeed, Ohashi et al.\cite{ohashi2023early} reported that C$^{18}$O (2-1) emissions were detected in all 19 protostars of the eDisk sample, and the interpretation that some emissions have characteristics of an equatorial outflow facilitates the understanding of the observations. For example, IRAS 04302 is a Class I protostar, also nicknamed "Butterfly Star", located within the L1536 cloud of the Taurus star-forming region, which has been observed in detail as part of this program. The gaseous emission of CO and its isotopes in the disk of IRAS 04302 is seen almost edge-on from Earth, and the equatorial outflow appears as the wings of a butterfly emerging from the protostar surrounding the conical space extending from 300 au to 640 au that is empty because the MRF remnants are moving away from the protostar and only become observable again beyond 640 au (see the image of the $^{13}$CO emission line in Figure 3 in Lin et al., \cite{lin2023early}). According to the RDS-MRF model (Lin et al., offered a different interpretation), we would be observing the formation of a ring of gas with the remains of the MRF beyond 640 au.
   \end{itemize} 
   
\begin{figure}[ht]
\centering
\includegraphics[width=10 cm]{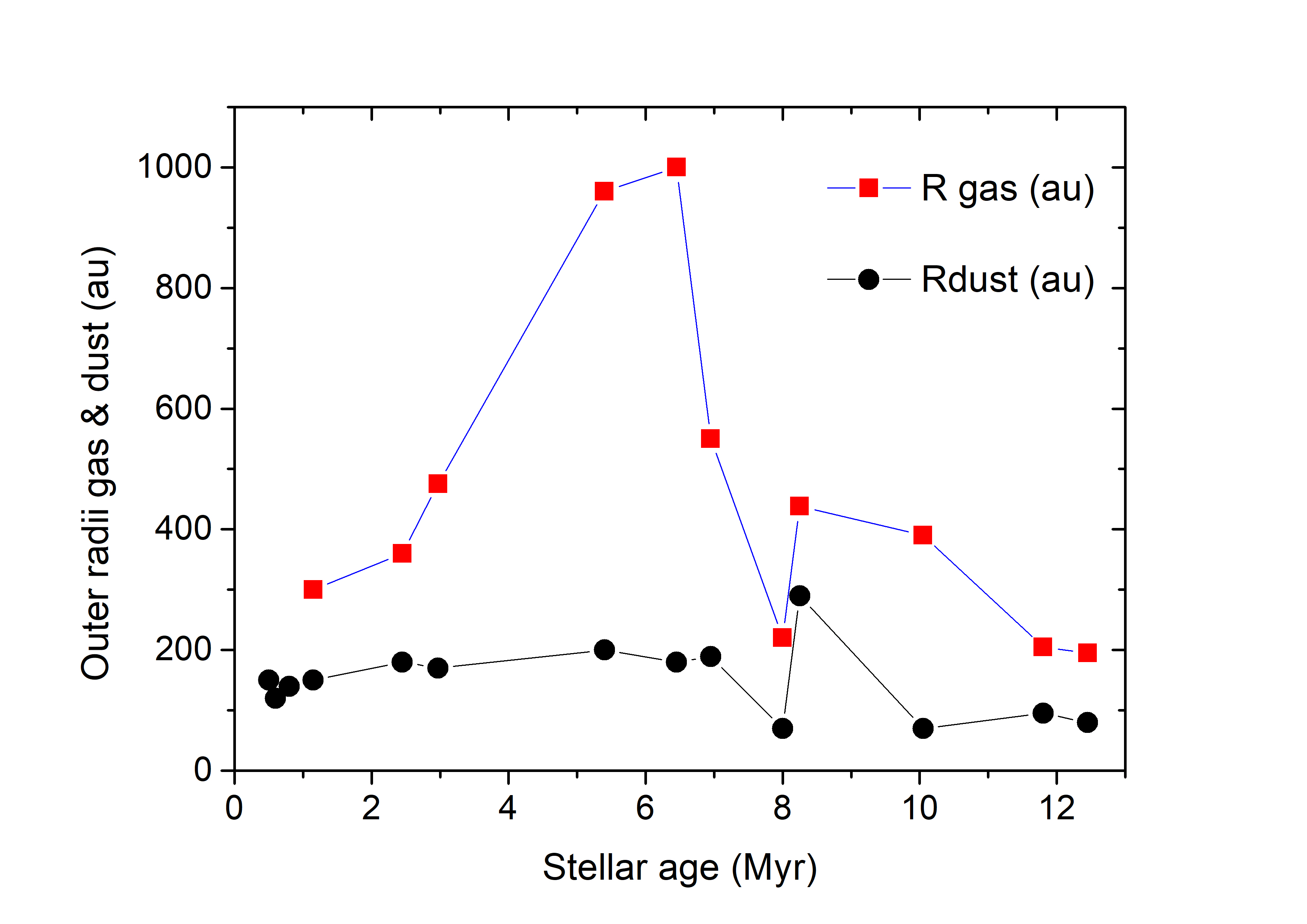}
\caption{Radii of the outermost gas (11 square dots) and dust rings (14 circular dots) of some protostellar disks as a function of stellar age. From youngest to oldest, the data obtained from van der Marel et al.,\cite{van2019protoplanetary} correspond to: Elías 24, HL Tau, GY 91, AS 209, Sz 98, RXJ1615, HD 97048, DM Tau, HD 163296, TW Hya, V1094 Sco, HD 100546, HD 135344B, HD 169142. The ages of stars often vary by a factor of 2 or more between different studies, so their average age has been calculated. Data for AA Tau and V1247 Ori are not represented because only the lower limit of R$_{gas}$ ($>$ 220 and $>$ 280 au respectively) has been reported for these disks. For the three youngest stars (HL Tau, Elias 24 and GY 91) there is no gas ring beyond the outer dust ring and there are only 11 R$_{gas}$ data in the figure.}
\label{fig:Nt1-5}
\end{figure}

\subsubsection{Observation of the equatorial outflow debris in the protoplanetary disks of class II protostars}

However, the best observation of the MRF has been made in the protoplanetary disks that surround the Class I and Class II protostars, where the remains of the MRF can be perceived as a ring of gas located beyond the edge of the dust disk from which it is separated by a nearly empty gap. A few years ago, van der Marel et al.\cite{van2019protoplanetary} compared the morphology and gap location of 16 protoplanetary disks using published ALMA data. One of the most remarkable findings is an equatorial gas ring  beyond the edge of the dust disk observed in all disks with few exceptions. Figure \ref{fig:Nt1-5} shows the radii of the outermost rings of gas (11 square dots) and dust (14 circular dots) of the protoplanetary disks as a function of stellar age. The gas ring that extends beyond the dust disk in 11 of the 14 samples studied has the characteristics of an equatorial outflow that continues to move away after have traversed the dust disk.

The interpretation offered by the RDS-MRF model is that the absence of a gas ring in the three youngest stars (Elías 24, HL Tau, GY 91) must be attributed to the fact that the equatorial outflow is still traveling through the disk and has not had time to reach its outer edge and separate from it. This interpretation also anticipates that after thousands of years each of these three equatorial outflows will reach the edge of their respective disk, creating a gap as they move away from it as occurred in the other eleven disks.

\subsection{The composition of the equatorial outflow (MRF): molecular ions in disks}

\begin{figure}[ht]
\centering
\includegraphics[width=12 cm]{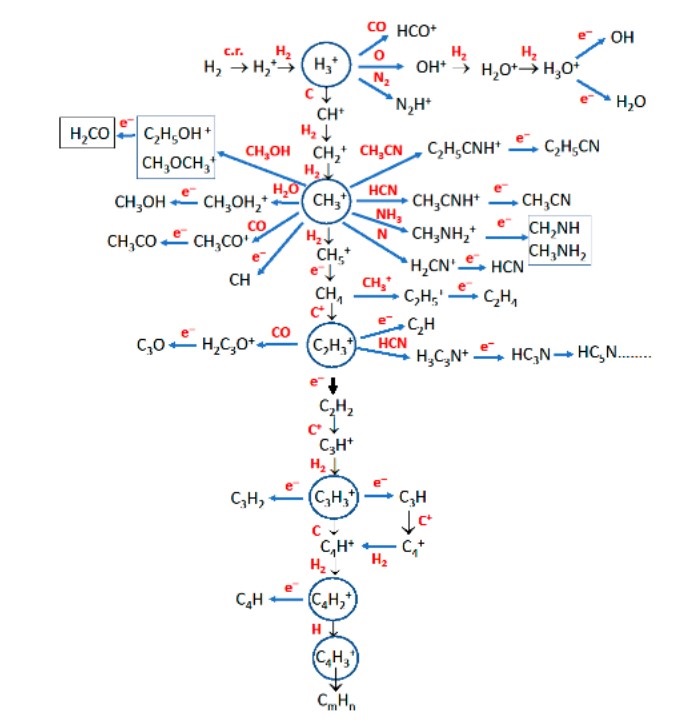}
\caption{Chemical reactions initiated by H$_3^+$ as a function of the composition of the medium that surrounds it (in red). This figure is based on the one presented by McCall in his PhD Thesis at the University of Chicago (2002) \cite{mccall2002spectroscopy}}
\label{fig:NT1-6}
\end{figure}

The MRF is a transient phenomenon and its composition at the time of emission is difficult to establish, and it is even more difficult to know how it evolves over time since it must change as it exhausts its most reactive components. However, the results obtained by the MAPS Large Program could give us some information about its composition.

Aikawa et al., investigated the ionization structure of the five disks of the MAPS program using the emission of HCO$^+$ and its deuterated isotopologues \cite{aikawa2021molecules}. Using theoretical models, they propose that the main molecular ions in disks are {H$_3$}$^+$, HCO$^+$, N$_2$H$^+$ and their deuterated isotopologues (see Figure \ref{fig:Nt1-2}). The abundance of atomic ions such as H$_3^+$, C$^+$, S$^+$ on the surface of the disk is generally attributed to photo-ionization and cosmic-ray ionization \cite{bosman2021molecules}, while molecular ions such as N$_2$H$^+$, N$_2$D$^+$, and H$_2$D$^+$ would predominate in the mid plane of the disk less accessible to radiation. The detection of H$_3^+$ in interstellar space has been carried out using rotation-vibration transitions and, since it cannot be detected at millimeter wavelengths it is not observed with ALMA. However, Aikawa et al \cite{aikawa2021molecules} include H$_3^+$ in the composition of both the mid plane of the disk and its atmosphere, since it is one of the most abundant interstellar molecules (only surpassed by H$_2$) and its presence is ubiquitous. 

The H$_3^+$ ion is one of the most important molecules in existence as it is the precursor to many chemical reactions, including those leading to compounds such as formyl cation (HCO$^+$), hydroxide ions (OH$^-$) and water, diazenylium (N$_2$H$^+$), and deuterated isotopologue of H$_3^+$, H$_2$D$^+$. The interstellar H$_3^+$ is produced (Eq. 6) by the ionization of H$_2$ by cosmic rays (cr). This leads to H$_2^+$, which promptly reacts with the abundant H$_2$ in the medium to give H$_3^+$ (Eq. 7).
\begin{equation}
H_2 + cr {\rightarrow} H_2^+ + e^{-}
\end{equation}
\begin{equation}
H_2^+ + H_2 {\rightarrow} H_3^+ + H
\end{equation}
The production of H$_3^+$ is balanced by the destruction reactions 
\begin{equation}
H_3^+ + X {\rightarrow} HX^+ + H_2 
\end{equation}
where H$_3^+$ acts as a universal protonator and X is an atom or molecule. HX$^+$ combines with other species and initiates a chain of reactions that produce new molecules; H$_3^+$ is also destroyed in dissociative recombination with electrons \cite{mccall2003enhanced,larsson2008h3+}.

However, for the H$_3^+$ ion to be found in the MRF, it must first exist in the protostar, and it seems that it does. Aoyama and Ikoma reported that when stars form, there is another possibility to ionise hydrogen besides cosmic rays \cite{aoyama2019constraining}. During the stellar accretion process, the high free-fall velocity causes a strong shock at the stellar surface, heating the accreted gas (mostly hydrogen and helium) up to temperatures at which the hydrogen is completely ionised. After Aoyama and Ikoma`s work, it seems possible that the composition of the equatorial outflow emitted by the protostar contains an abundance of H$_3^+$ ions. Due to its great reactivity and acidity, the H$_3^+$ ion would play an essential role in triggering chemical reactions in a medium such as the disk rich in molecular hydrogen, carbon molecules and ions, and electrons.

As is illustrated in Figure 6, all the main molecular targets of MAPS, except CO and CN, can be obtained from chemical reactions that are triggered by the arrival of H$_3^+$ to a disk rich in H$_2$, HD, O$_2$, CS, CO, CN, SO, etc. \cite{mcguire20182018} This would explain the rapid formation of the new molecules that launch the nucleation process in the disk. Moreover, since the density at the disk surface is much lower than inside the disk, the number of chemical reactions in that region must also be lower but not negligible. On the other hand, CO and CN are abundant molecules in the ISM and must also be abundant in the protostellar disks that form around Class 0 and Class I stars. It seems that CO could be a component of both the disk and the MRF.

In the RDS-MRF example shown in Figure \ref{fig:Nt1-1} B there are only two reactants in the system. If we consider the protostar and its protostellar disk, the number of reactants should be considerably higher due to the complexity of the composition and the size of the system. However, it could be the case that there are not as many reactants as the size of the system seems to suggest, and that the protostar-disk system remarkably resembles a two-reactant RDS-MRF. If the H$_3^+$ ion was part of the MRF the chain of molecular ions that are synthesised from it on its way through the disk (CH${_3}^+$, C$_2$H$_3^+$, …, C$_m$H$_n^+$) could play a role analogous to that played by the phosphate ions in the MRF of the example shown in Figure \ref{fig:Nt1-1} B. The resemblance between the disk-protostar system and the two-reactant system in Figure \ref{fig:Nt1-1} B may be greater than expected.

\subsection{Particle rings analogous to nucleation fronts}

The search for the Nucleation Front involves searching for the regions of the protoplanetary disk where micrometric particles are formed. The size of the dust grains in molecular clouds ranges from about 100 nm to 500 nm, with a peak at 220 nm generally attributed to small carbonaceous particles \cite{Williams2005}. Within the protostar, the size of the dust grains will likely decrease as part of their coating evaporates and they will not be detectable in ALMA observations. In contrast, in the disks the 870 $\mu$m and 1.3 mm continuum images obtained with DSHARP \cite{huang2018disk,dullemond2018disk} provide information on the spatial distribution of nuclei, which the RDS-MRF model predicts will appear in the disk after the passage of the equatorial outflow. ALMA observations (see bottom panel in Figure \ref{fig:Nt1-3}) show that Class II disks have a multi-ring and gap substructure in both dust and gas distribution. The mass stored in each of these rings varies between one Earth mass or even less, and several tens of Earth masses. The radial brightness distribution of the dust rings can be modeled with the sum of concentric Gaussian rings \cite{guzman2018disk}, setting r=0 at the center of the innermost Gaussian, i.e., at the center of the interface, 
\begin{equation}
I(r) = \sum_{i=0}^N  A_i exp \frac {-(r-r_i)^2} {2 {\sigma_i}^2}
\label{Eq:Nt1-11}
\end{equation}
indicating that the micrometer/millimetre dust grains are located in radially confined rings where migration between rings has stopped and, most importantly, the intensity of the peaks is maximum in the centre of the ring. These dust rings, which in RDS-MRF terminology would correspond to the NF, have been called “dust traps” since they provide an ideal environment in which radial drift is not important and grains can grow to form aggregates and boulders.  

The RDS-MRF model predicts that particles that form earlier within the rings closer to the protostar will be larger because they have more time to grow than particles in the outermost rings that form later. The appearance of gaps between the particle rings is due to the time lag between the passage of the MRF and the beginning of nucleation, not to the presence of planets. The model predicts that planets will form in the regions where particles accumulate and that occurs in the central part of dust rings, where particles can pile up to several Earth masses. The position of the maximum of the Gaussian fit is at the center of the rings, indicating that this is where the particles end up concentrated and where the planets will form. 

The RDS-MRF model is only valid for an MRF emitted by a single source and would not be applicable in binary or ternary systems in which the emission from several sources alters the formation of the respective disks and the structure of the particle rings. Binary and multiple systems not only show specific behaviors that are difficult to summarize but also pose the problem that in literature there are no RDS-MRF models to compare them with. The structure of binary protostellar systems is very rich and varied given the wide range of combinations of variables that can arise in this case. For example, the multiple system UX Tauri shows large spirals forming a bridge between UX Tau A and UX Tau C. The disk of UX Tau C is much smaller than the disk of UX Tau A and they are possibly misaligned \cite{menard2020ongoing}. Sometimes the curious situation arises that binary systems were formed asynchronously, so the formation of their disks and the emission of their respective MRFs would not be simultaneous.

\section{Discussion: Disk structure as a function of the time it takes the MRF to traverse it} \label{sec:Evolution}

The RDS-MRF model allows us to explain the evolution of the disk surrounding single stars as a function of the time it takes the equatorial outflow to traverse the disk and, once it reaches its outer edge, to move away from it until the star's gravity balances the impulse with which the MRF was launched.
Since the equatorial outflow is launched from the star, the disk structure evolves from the inside out, and to apply the scaling laws of the RDS-MRF model, the protostar/disk interface must be taken as the origin of coordinates and the instant of the MRF launch as the time origin. According to the spacing law, the first particle rings are very close together, and their separation increases with distance from the star. The width law predicts that the first particle rings are narrow, and their width increases with distance from the star. Finally, the time law predicts that particle rings form from the inside out.

The RDS-MRF model explains the evolution of the disks from a continuous structure to a ring/gap structure, resulting from the time lag between the MRF and the NF. The beginning of planetary accretion in the first particle rings is visualized in ALMA as a cavity surrounding the protostar. The radius of this cavity increases as the disk disappears with the formation of the outer planets, eventually leaving only the outermost rings, in which, due to their great width, particle accretion is hindered or ceases completely. Much farther from the star lies a debris ring generated by the remnants of the MRF. In the Solar System, this disk would correspond to the Hills cloud, located between 1,000 and 10,000 au from the Sun, whose existence is inferred from the distribution of the semi-major axes of long-period comets that reach the inner Solar System with low-inclination orbits \cite{vokrouhlicky2019origin}.

The RDS-MRF model distinguishes three different causes that can give rise to regions in the disk depleted of dust particles, namely: i) the delay of nucleation with respect to the passage of the MRF creates gaps between the particle rings when gravity acts, aggregating the particles that have formed, ii) the formation of planetesimals and planets creates a huge gap (cavity) in the region surrounding the star, and iii) the motion of the MRF leaves a big gap as it moves away from the dust disk. By distinguishing between three different causes of gaps in the disk, the RDS-MRF model naturally explains the transformation of continuous disks surrounding very young single stars into ring-shaped disks and subsequently  in transition disks, as a cavities are created around the star due to the formation of planets by the accretion of the particles in the rings that according to the time law, first appear in the region near the star.

\begin{figure}[ht]
\centering
\includegraphics[width=\linewidth]{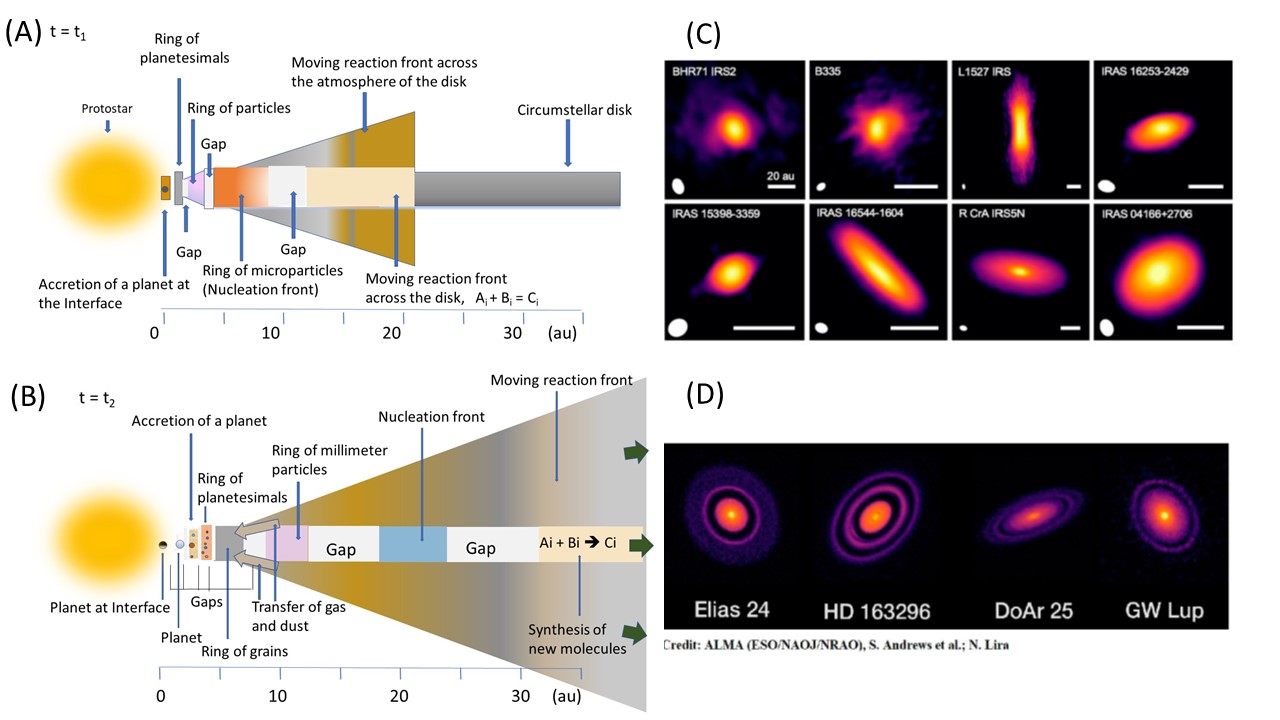}
\caption{Early disk structure. (A) Continuous disks: the initial structure of the disk around single young stars is smooth. View at t = t$_1$ of the protostar and the protostellar disk seen edge-on as it is traversed by the MRF. In the disk closest to the protostar, a ring of nuclei has evolved into boulders and planetesimals, a little further from the star and separated by a gap would be a ring of aggregates and particles. As we move away from the star, the nucleation front would be observed as a ring of micro particles. Further away we would find the region where the matter of the disk reacts with the matter transported by the MRF and finally the part of the disk that the MRF has not yet reached. At a resolution of approximately 5 au, the disk will appear continuous. (B) Disks with annular substructure: at time t$_2$ $>$ t$_1$ the MRF reaches the outer edge of the disk and continues moving away from the protostar, creating a gap between the edge of the disk and the remains of the MRF. Meanwhile, in the part of the disk closest to the star, the particles under gravity have accreted into planetesimals and planets. The arrows point to a possible mechanism for gas transfer from the disk to the particle ring \cite{francis2020dust}, which would facilitate the formation of a subnebula at the position indicated by the spacing law. Note that in this figure panels (A) and (B) are not drawn to scale. (C) Eight continuous Class 0 disks observed  within the eDisk program and reprinted with permission from  Ohashi \cite{ohashi2023early} and Lin \cite{lin2023early}. (D) Four Class II disks observed within the DSHARP program  showing a compact inner region and spaced rings on the outermost part of the disk. Credit: ALMA(ESO/NAOJ/NRAO), S. Andrews et al.; N. Lira.}
\label{fig:Nt1-7}
\end{figure}

\subsection{Continuous disks in young stars}

  At the time of their formation, t = t$_1$,  the structure of the disks is continuous. It is the time it takes for the MRF to travel the disk that defines whether the disk is going to be continuous or ringed. It is the time it takes the nuclei to form at the different regions on the disk, that decides whether micro-metric or millimeter particles will be observed. It is the time it takes for aggregates and planetesimals to form near the protostar that will decide whether the disk has a cavity or not. For example, in the case of very young stars it is possible that at the time of observation the equatorial outflow is still traveling along the disk without having reached its edge, in which case the rings may not have formed yet or are in the process of doing so. The rings near the star are the first to form (time law), are very narrow (width law) and close to each other (spacing law). It is possible that even if one or two inner rings have formed, they cannot be seen with ALMA (resolution $\sim$ 5 au at 125 pc) and the disk would look continuous (see the drawing in Figure \ref{fig:Nt1-7} (A) and the observations of continuous disks with ALMA in \ref{fig:Nt1-7} (C). 

One example of continuous disks appears in the study carried out by Long et al. \cite{long2019compact} in ALMA with high resolution ($\sim$ 0".12 $\sim$ 16 au) and average sensitivity (50 $\mu$Jy beam$^{-1}$ at 225 GHz) of 32 protoplanetary disks around stars of solar mass in the star-forming region of Taurus. Of these disks, 12 around young stars (between 0.9 and 5.9 Myr old) appear as smooth disks lacking substructure. In this study, the median age for smooth disks is 2 Myr whereas for disks with ring/gap substructure is 3.2 Myr. Recently, the eDisk program made observations of twelve nearby Class 0 disks ($<$ 120 pc) and found that none of them have a ring/gap substructure nor is a ring of gas observed beyond the dust disk \cite{ohashi2023early}. No rings are seen in the disks of Class 0 and we do not know if this is due to the 5 au average resolution of ALMA, or because the rings close to the star have not yet formed. The only case in which the resolution of 1 au has been achieved with ALMA was presented by Andrews et al. \cite{Andrews_2016} observing the disk around TW Hya. The high-resolution image reveals a series of narrow, bright concentric rings separated by narrow dark gaps in the region between 1 and 6 au, fulfilling both the spacing and width laws, which predict that wide rings and gaps should appear only at large distances from the protostar.

\subsection{Disks with annular substructure}
It was unexpected to observe that many disks, instead of a smooth and continuous structure of dust and gas, have a well-defined structure of rings of particles separated by empty gaps. It would seem that there can be no common mechanism that relates the structure of the images obtained in the eDisk program (Figure \ref{fig:Nt1-7} C) to the images of the Class II disks (Figure \ref{fig:Nt1-7} D), and yet the RDS-MRF model provides a narrative that connects the two through the moving reaction front.
After the passage of the equatorial outflow, at time t$_2$ $>$ t$_1$, the continuous disks develop an annular substructure of rings of particles separated by gaps. In the scheme of Figure \ref{fig:Nt1-7} (B), the MRF has already reached the disk edge and is beginning to move away from it. The part of the disk through which the MRF has just passed is an area of great chemical activity, while as we approach the protostar we find rings of nano- and microparticles corresponding to the nucleation front. Rings delimit the regions where particles will grow and planets will emerge while gaps separate the feeding zones of each planet.
The possibility that planet formation creates the gaps between the rings of disks has often been raised, although no direct images of planets in these gaps have been obtained. However, one place where planets might be found is in the cavities near some stars observed with ALMA in so-called transition disks \cite{espaillat2007diversity,francis2020dust}.

The arguments for treating the protostellar disk system as an RDS-MRF are numerous. According to this model, the rings of micro- and nanoparticles observed in the disks of Class I and II protostars can be considered the equivalent of nucleation fronts. With current data, it seems that the scaling laws of RDS-MRFs describe quite well the disks structure. Furthermore, the absence of a gas ring beyond the dust disk in only the three youngest protostars shown in Figure \ref{fig:Nt1-5} supports the interpretation that the MRF is still traveling through the disk and has not yet reached its outer edge.

\begin{figure}[ht]
\centering
\includegraphics[width=15 cm]{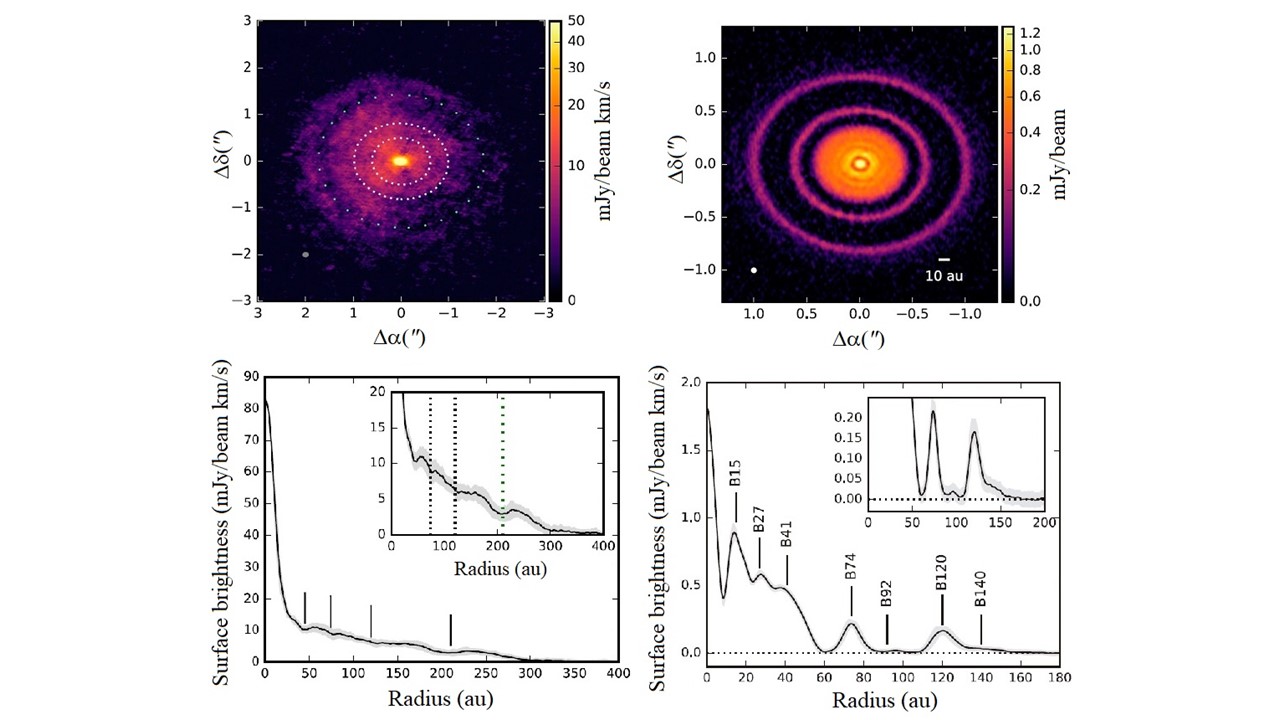}
\caption{Gas and dust disks around AS 209. Moment-zero map (top left panel) and azimuthally averaged radial profile (bottom left panel) of the $^{12}$CO (2-1) rotational line emission. The two white dotted lines in the top left panel mark the position of the two outer dust rings located at 74 and 120 au, while the green dotted line illustrates the position of the outermost CO gap at about 220 au.  The four vertical lines in the bottom left panel mark the position of the CO gaps at radii near 45, 75, 120 and 210 au. The right side of the Figure shows the dust continuum emission map (top right panel) and the azimuthally averaged radial profile of the dust emission (bottom right panel) of the disk. The seven vertical lines in the bottom right panel mark the position of the center of the dust rings (B stand for brightness). Insets are included in the radial profile panels to show the faint peaks. The figure is reprinted from Guzman et al.\cite{guzman2018disk} and reproduced with her permission. Copyright AAS}
\label{fig:Nt1-8}
\end{figure}

Next, the disk around the protostar AS 209 will be analyzed in detail. This disk has been selected because it shows an example of annular structure, and also because it has been studied in depth by several groups \cite{huang2018disk,fedele2018alma,guzman2018disk} and the results reported by all of them agree well. The details of the disk substructure shown in Figure \ref{fig:Nt1-8} have been taken from the DSHARP project in which the resolution achieved was 0".037, corresponding $\sim$ 5 au. These images have been published and analyzed in detail by Guzman et al., \cite{guzman2018disk} where they report all the information related to data acquisition, applied corrections, image production and an interpretation different from that offered by the RDS-MRF model.

 The upper part of Figure \ref{fig:Nt1-8} illustrates the spatial distribution of $^{12}$CO (top left panel) and the 1.25 mm dust continuum emission (top right panel) around AS 209, while the respective azimuthally averaged radial profiles are shown in the lower panels.  In the RDS-MRF model, the distance to the star of the exhausted MRF is greater than that of the last ring of particles and the same is true in AS 209 where the $^{12}$CO emission extends much farther out (peak at $\sim$ 240 au) than the millimeter dust emission (last peak at $\sim$ 140 au). The radial profile of the $^{12}$CO emission would indicate that the MRF is currently located about 240 au from the star, and its edge extends out to 300 au (Figure \ref{fig:Nt1-8}, lower left panel).
 
The dust emission of AS 209 consists of two well-differentiated regions, a group of narrow and very compact rings close to the star ($<$ 60 au), and two bright and well-separated rings at 74 y 120 au (Fig. \ref{fig:Nt1-8} right side). The bands of particles in the RDS-MRF model are more compact close to the interface than in the region far from it because they obey the spacing law (r$_n$/r$_{n-1}$ $\rightarrow$ p $>$ 1). Let's imagine that r$_{n-1}$ = 1 and p = 2, then the bands will appear in positions 1, 2, 4, 8, 16, 32, 64, etc. With a resolution of 5, the first three bands will appear as a compact region, while the two most distant bands will appear well separated. Thus, the existence of two well-differentiated regions in ADS 209 (see Figure \ref{fig:Nt1-8} upper right panel) and in many other protoplanetary disks such as DoAr 25 (Figure \ref{fig:Nt1-7}D), Elias 24 \cite{cieza2017alma}, HD163296 \cite{isella2016ringed} and Gw Lup \cite{huang2018disk} can be explained by the spacing law. 

In the disk of AS 209 the position of the dust rings is 15, 27, 41, 74 and 120 au \cite{guzman2018disk}. Using these values we obtain p = 1.8; 1.5; 1.8; and 1.62 (the position of the last ring according to Fedele et al. \cite{fedele2018alma} is 130 au, this would give p = 1.8 instead of 1.62), which seems to indicate a spacing law with p $\rightarrow$ 1.7 or 1.8. Furthermore, there is evidence that a spacing law can be applied to the more than 4,000 exoplanets that have been found so far, which follow the same spacing law as the particle bands where they grew \cite{bovaird2013exoplanet}. As for the time and width laws, there is no information on the time of formation of the rings, nor on their width, so they cannot be discussed here, although the large width (about 7 - 10 au) of the two outer rings of AS 209 supports the width law \cite{guzman2018disk}.

It is surprising that the RDS-MRF model works despite not taking into account either the rotation of the protostar or the disk, which seems to indicate that the MRF is emitted in the early stages of the collapse of the nebula when the angular velocity difference between the protostar and the disk was small and the system had a high degree of symmetry.

\subsection{Transition disks}
 There are also cases where the disks exhibit large holes in regions close to the star, the so called cavities. Transition disks, which according to the RDS-MRF model appear at time t$_3$ $>$ t$_2$, are characterized by having au-scale  cavities in their dust distribution near the star, which causes a deficit in infrared emission in their spectral energy distribution \cite{muzerolle2009spitzer,mauco2020naco}. According to Muzerolle at al., \cite{muzerolle2009spitzer} there are signs of age dependence manifested in the lack of transition disks in the youngest stars, while their frequency increases for ages 3 to 10 million years. The most accepted hypothesis attributes the cavity to the presence of one or more planets that clean the disk in the star´s surrounding region

Transition disks are recognized by a reduced near-infrared excess compared to the median SED of disks of young stars, which may indicate that they have developed some radial structure. This NIR reduction is attributed to the disk dissipation from the inside out \cite{williams2011protoplanetary,hashimoto2015structure} which generates a deficit in the SED of the object to $\sim$ 10 $\mu$m. Their inner regions are devoid of dust grains in a radius of up to several tens of astronomical units whereas the outer regions still contain substantial amount of dust. According to Wyatt \cite{wyatt2008evolution}, a new paradigm is emerging around the idea that disks are cleaned from the inside out and this cleaning is increasingly attributed to planet formation rather than photo evaporation \cite{clarke2001dispersal}.

\begin{figure}[ht]
\centering
\includegraphics[width=15 cm]{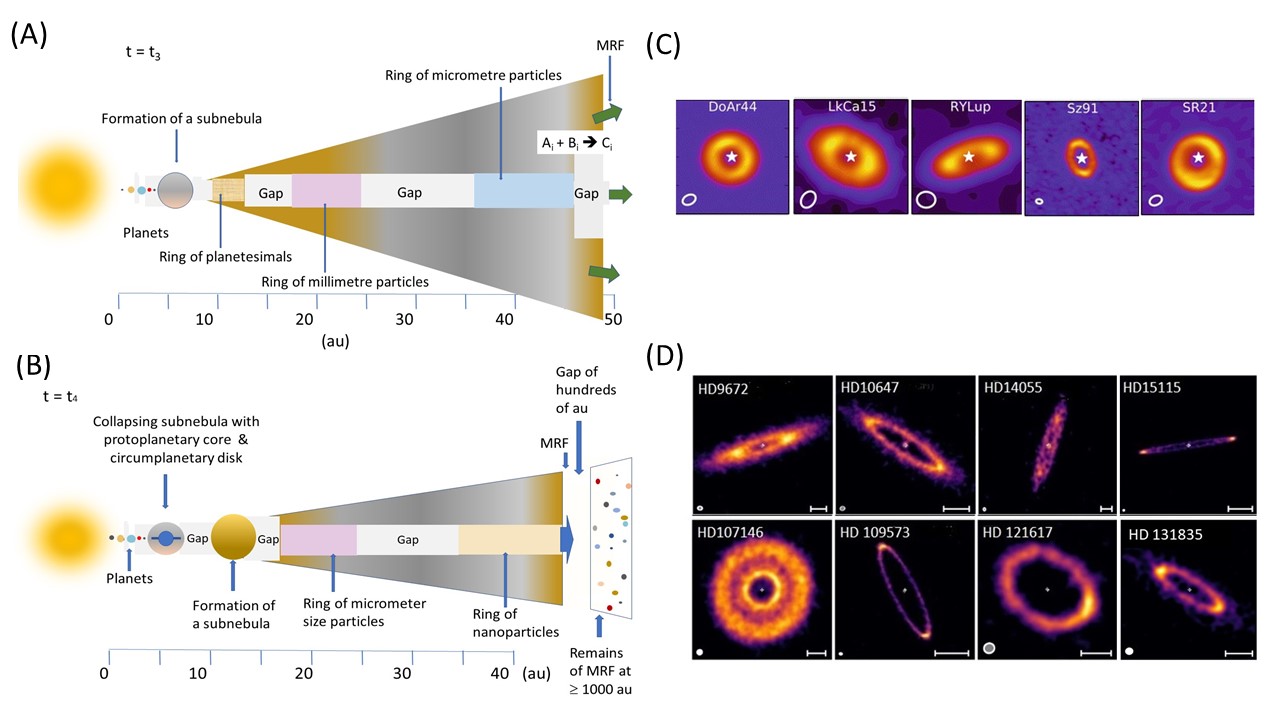}
\caption{(A) Drawing of the cross section of a transitional disk at time t$_3$ $>$  t$_2$ of its evolution if it would behave like an RDS-MRF. In the part closest to the star, planets and planetesimals have already formed, while in the more distant regions the accretion process has not yet begun and rings of particles are observed. With ALMA, the disk would look like a transition disk with a radius cavity about 8 au, a planetesimal ring at 10 au and two particle rings at 20 au and 40 au. Separated by a gap from the outer edge of the disk, around 50 au a ring of gas corresponding to the MRF would be observed moving away from the disk. (B) Sketch of the cross-section of a debris disk at time t$_4$ $>$ t$_3$ with two particle rings at 20 and 40 au from the star and a ring much farther from the star with the remains of the MRF. (C) ALMA images of five transition disks showing a cavity around the star. Credit: ALMA(ESO/NAOJ/NRAO)/NASA/ESA. (D) ALMA images from the ARKS program of eight debris disks showing a halo at distances from the star ranging from 30 au to 220 au. Credit: Marino et al.\cite{marino2026alma}}
\label{fig:Nt1-9}
\end{figure}

Several scenarios have been proposed to explain the substructure of transition disks, such as disk-planet interaction \cite{dodson2011transitional}, photo evaporation \cite{clarke2001dispersal}, grain growth \cite{birnstiel2012can}, and a few others. The RDS-MRF model points out that the origin of the cavities in the transition disks is the formation of planets and planetesimals that according to the time law precedes that of the planets furthest from the protostar (although the time law refers to the formation of particle rings, the formation of planets is assumed to follow it). Even though  the cleaning of the internal regions of the disk should start very early (Figure \ref{fig:Nt1-7} B) can only be observed with ALMA when the cavity radius exceeds about 5 au. At time t$_3$, in the part of the disk close to the star, planets would have already formed and others would be forming, while increasing the radius we would still find particle rings (Figure \ref{fig:Nt1-9} A).
 
A little over a decade ago, Williams \& Cieza \cite{williams2011protoplanetary} and Armitage \cite{armitage2011dynamics} pointed out the possible presence of planets within the inner cavities of transitional disks, and since then, efforts have focused on detecting Jupiter-mass planets that could have created these cavities. However, the RDS-MRF model indicates that the planets close to the star (radius $<$ 5 au) must be small (width law) and this is why they are difficult to observe. This prediction was confirmed when ALMA reached 1 au resolution by observing TW Hya (56 $\pm$ 7 pc distance) and narrow rings of particles separated by narrow gaps appeared in the region between 1 and 6 au around the star \cite{Andrews_2016}.  

Giant planets will only appear when the cavity radius is greater than $\sim$ 5 au. In fact, two giant planets have been detected within the transition disk cavity of the star PDS 70: the protoplanet PDS 70b at $\sim$ 22 au, and shortly after a second protoplanet PDS 70c was located at $\sim$ 34 au from the star \cite{haffert2019two,aoyama2019constraining,isella2019detection}. 
Planets have also been detected within the LkCa 15 transition disk cavity where Sallum et al. \cite{sallum2015accreting} detected the presence of three protoplanets LkCa 15b, and LkCa 15c in Keplerian orbits with 14.7 au and 18.6 au semi-major axis respectively, and a third planet, LkCa 15d, which is considerably fainter than the others. The detection of planets around PDS 70 and LkCa 15 are the most robust observation to date on the formation of new planets still embedded in their natal disk, and occur in the transition disk cavity near the star. Related to this point arises the prediction of the RDS-MRF model that the size of the inner cavity should increase from the inside out as the star ages, extending throughout the disk until it eventually disappears. Thus, the evolution from transition disks to debris disks is gradual rather than abrupt, as the disk disappears from the inside out as the giant planets form. The dust sometimes found in the cavities of transition disks, according to the RDS-MRF model, is associated with the formation of satellite systems around the giant planets.

At this stage of evolution the gaps in the disk have three different origins. Namely: the formation of planets and planetesimals create the cavity near the star; the time lag between the passage of the equatorial outflow and nucleation creates the gaps between the particle rings in the disk; the motion of the equatorial outflow away from the disk creates the gap that separates it from the edge of the dust disk.

\subsection{Debris disks}
The disks surrounding young stars change their structure and composition over time, with the debris disks (DDs) that appear for t$_4$ > t$_{3}$ being the last stage of their evolution. Debris disks are gas-poor disks around main sequence stars that are characterized by having a SED that can be approximated by a black body at a single temperature. According to Wyatt \cite{wyatt2008evolution}, the debris disks can be defined by two parameters: temperature, T, and fractional luminosity, f, which is the ratio of the infrared luminosity of the dust (L$_{IR,dust}$)  to that of the star (L$_*$),

\begin{equation}
    f = \frac{L_{IR,dust}}{L_{\star}}
\end{equation}

these two parameters can be estimated from the wavelength $(\lambda)$ and flux (F$_{\nu}$) at the maxima in the emission spectra of the disk and the star.

\begin{equation}
    T = \frac{5100}{\lambda_{disk,max}}
\end{equation}
\begin{equation}
    f = \left( \frac{F_{\nu disk, max}}{F_{\nu \star, max}} \right) \left( \frac{\lambda_{\star, max}}{\lambda_{disk, max}} \right) 
\end{equation}

where $\lambda$ is in micrometers and T in Kelvins. The defining property of a debris disk is that it has a fractional luminosity f $<$ 10$^{-2}$, lower than that of protoplanetary disks, since its temperature can vary between 10 and several 100 K.

The abundance and properties of debris disk (DDs) around stars of different age, mass and luminosity has been addressed in different reviews \cite{hernandez2007spitzer,wyatt2008evolution,matthews2014observations,hughes2018debris}. Debris disks are made up of mainly solid bodies whose size varies from a few microns (dust) to thousands of kilometers (planetesimals and planets). The small component of dust and gas in DDs can be detected by absorption or emission. It is believed, that the dust is continuously regenerated trough collisional grinding of previously formed solid bodies, what constitutes the main characteristic of these disks. Another feature that serves to differentiate them from transition disks is that in the latter the dust is optically thick at optical wavelengths, while in the debris disks it is optically thin for the entire electromagnetic spectrum. The Solar System seen from a distance of tens of parsecs would appear as a debris disk with nine planets and two dust rings: a zodiacal ring in the terrestrial zone and a second ring in the region of the Kuiper Belt. Zodiacal dust has orbital properties coinciding with those of some asteroid families, suggesting that the dust originates from collisions between asteroids. Likewise, the dust of the Kuiper Belt would have its origin in the collisions between bodies present in that region. It is still debated whether the dust and gas observed in debris disks are remnants of what existed in their protoplanetary disk (called primordial) or were created at a later stage (called secondary), although all indications point to it being secondary, as is believed to occur in the Solar System.
	
Near infrared emission decreases from $\sim$ 100 \% to 0 \% over 6 to 10 Myr in stars similar to the Sun, which seems to indicate that this is the average age of the protoplanetary disks associated with them \cite{hernandez2007spitzer}. The fact that the main disk disappears does not mean that the planetary formation process has been completed since at that time, although the planets close to the star (terrestrial planets) may have already formed in the outermost regions there would be giant planets in process of forming their respective satellite systems (see Figure \ref{fig:Nt1-9} B). From the moment of their birth debris disks undergo a considerable evolutionary process until they reach several billion years. It could happen that primordial dust associated with the planetary subnebula is observed in young debris disks while in older disks only secondary dust is observed as in the Solar System.

Very recently, the ARKS program uses ALMA to provide high-resolution images of a sample of 24 debris disks \cite{marino2026alma,han2026alma,jankovic2026alma} many of which show wide dust belts in the outer regions (see Figure \ref{fig:Nt1-9} D) rather than narrow rings.
The RDS-MRF model distinguishes between disks (i.e., HD145560, HD 61005 and HD121617...) \cite{han2026alma} that show wide rings centered between 60-70 au that would confirm the validity of the width law and indicate the presence of planetesimal belts similar to the Kuiper Belt, while other disks (i.e., HD15257, HD84870, HD14055, HD76582, HD95086, HD218396, HD170773...)\cite{han2026alma} show much wider belts at distances $\approx$ 200 au that would correspond to MRF debris moving away from the star. In some other disks (i.e., HD107146, HD92945, HD206893...)\cite{han2026alma} an intermediate structure would be observed with wide rings at distances  $\approx$ 50 au and very wide belts at a distance $\geq$ 100-150 au that would correspond to exoKuiper belts and the debris of the MRF respectively. Due to the low density of planetesimals in these belts, the formation of large planets within them is scarce or does not occur.

\section{Conclusions} \label{sec:Conclusion}

The RDS-MRF model provides a narrative to describe the evolution from the continuous disks surrounding class 0 stars to the ring/gap structure in class II disks, and the subsequent formation of cavities in the transitional disks. These cavities, which extend from the inside out, along with the low density of planetesimals in the outermost belts, give rise to the formation of debris disks structure. The motion of the MRF imposes a temporal evolution that, following cause-and-effect relationships, qualitatively explains and predicts the disk categories observed with the ALMA radio telescope in the various programs developed over the past 10 years.

\subsection {Acknowledgements and competing interest}
ELC thanks: i) Chein-Shiu Kuo and A. Scala for the interesting discussions on reaction-diffusion systems and astronomy, ii) Prof. Carmen Rueda, Dr. Margarita Menéndez, and Dr. Juan Pedro Cascales for their constant support.

\subsection {Authors contribution}
This declaration is not applicable because this article is done by one author.

\subsection {Funding}
The author received no financial support for the preparation of this article.

\subsection {Competing interest}
The author declares that he has no competing interests that could have influenced the work presented in this article.

\subsection {Data availability}
The data used for the research described in this article come primarily from the programs running on ALMA, namely: DSHARP, MAPS, eDisk, FAUST, and ARKS. Permission to use the images was obtained from the corresponding author of the article where they appear. Full details on data and image acquisition and processing can be found in the references to these ALMA programs throughout the article. Figure 4A in the article is adapted from that presented by Greenhill et al.\cite{greenhill1998coexisting}, copyrighted by Springer Nature, and reused under license number 5667730238292.

\bibliography{sample}

\end{document}